# Making grains tangible: microtouch for microsound

Staas de Jong
LIACS, Leiden University
Niels Bohrweg 1, Leiden
staas@liacs.nl

## ABSTRACT

This paper proposes a new research direction for the large family of instrumental musical interfaces where sound is generated using digital granular synthesis, and where interaction and control involve the (fine) operation of stiff, flat contact surfaces.

First, within a historical context, a general absence of, and clear need for, tangible output that is dynamically instantiated by the grain-generating process *itself* is identified. Second, to fill this gap, a concrete general approach is proposed based on the careful construction of non-vibratory and vibratory force pulses, in a one-to-one relationship with sonic grains.

An informal pilot psychophysics experiment initiating the approach was conducted, which took into account the two main cases for applying forces to the human skin: perpendicular, and lateral. Initial results indicate that the force pulse approach can enable perceivably multidimensional, tangible display of the ongoing grain-generating process. Moreover, it was found that this can be made to meaningfully happen (in real time) in the same timescale of basic sonic grain generation. This is not a trivial property, and provides an important and positive fundament for further developing this type of enhanced display. It also leads to the exciting prospect of making arbitrary sonic grains actual physical manipulanda.

## 1. INTRODUCTION
## 1.1 Granular synthesis of musical sound, and its instrumental control

During the 19th and 20th centuries, newly developed technologies included increasingly practical methods to capture, transform and reproduce fragments of sound. When this is done for musical purposes, and the fragments involved have a brief duration of 0.1 s or less, the term *microsound* is often used [7]. In 1960, the composer Iannis Xenakis coined the term "grains of sound" in this context, also proposing a number of mathematically defined compositional tools for combining these grains into musical sound [10]. Currently, in granular synthesis a *grain* is defined as a sound fragment of duration 1 to 100 ms, resulting from a waveform signal shaped by an amplitude envelope. Over the years, composers increasingly have adopted granular techniques to create music, resulting in influential early works by Iannis Xenakis, Horacio Vaggione, Curtis Roads, Barry Truax, and others, and today granular sound synthesis is in widespread use.

Grain-based approaches to making musical sound were first implemented in a cumbersome process using analog magnetic tape technology. The subsequent revolution in the power and availability of digital computing technology, however, enabled the implementation and use of a series of increasingly sophisticated and powerful versions of granular sound synthesis [7]. It also enabled the introduction of implementations where the actions of instrumental control could occur simultaneously with the listening to their results [9]. Today, there are many such real-time implementations of granular sound synthesis available, often controlled using Graphical User Interfaces (GUIs) and the input from various types of MIDI controllers.

## 1.2 The interest of giving grains a dynamically instantiated tangible presence

One important use of the tangible aspects, in general, of instrumental control, in general, is display: to inform the human actions that are performed. Another important use is in defining *how* these actions can be performed, in the manipulations that are made possible. In the case of granular synthesis of musical sound, the object of such tangible display and manipulation will be the process of grain generation.

In existing systems, tangible display and manipulation are usually implemented using various types of general-purpose controller hardware, such as buttons, sliders, knobs, pads, and keys. These can then be used to initiate, modulate and terminate processes of grain generation in real time. However, the display and manipulation enabled by these components will not be very specific to the processes of grain generation that are controlled. Stages of tangible display and manipulation can be usefully set up to coincide with stages of grain generation. (E.g. as when overcoming the specific friction of moving a slider to a certain position, while this is being mapped to, say, the granular density.) However, this type of control is fundamentally limited, by the fact that it is not the process of grain generation itself that determines the tangible feedback.

In practice, these existing types of tangible display will give relatively little information about the grain generation in progress. As a consequence, for specific and detailed information, human operators will largely rely on the auditory display provided by the output of musical sound. This reliance has inherent disadvantages, e.g. in that the response of human actions to auditory feedback necessarily will be slower than the response to tangible feedback, making control less immediate [8].

For the above reasons, the existing real-time instrumental control of granular synthesis could be improved by using new forms of tangible display directly determined by the process of grain generation itself. These could provide the human operator with more information for her/his control, while this information could be made more specifically relevant; and could be delivered more immediately than is currently the case.

One such new form of tangible display is represented by the set of interfaces described in [4]. In these, the sound resulting from manipulations with colliding, breaking, deforming and sliding items is used to trigger and parametrize the digital generation of sound grains in real time. The stated main motivation for this approach was to expand, for musical purposes, the sonic range of familiar tangible manipulations. However, the resulting forms of instrumental control also have the property that the grain-generating processes directly determine tangible display, giving the advantages above.

What remains to be done, however, is to give these advantages, in the same way, to the widely used algorithmic processes of grain generation running on digital computing hardware in general. Here, the generation of each granular sound fragment will happen according to a set of explicitly defined parameters. Usually, the values for these parameters are uniquely determined for each grain, at the moment of its instantiation, to then remain fixed for the rest of its duration. Therefore, in order for tangible display to provide information that is as complete as possible, it should be capable of providing each grain instance with its own, dynamically determined tangible representation.

Having identified some necessary and desirable characteristics for new forms of tangible display for algorithmic grain generation, this can now motivate and guide the investigation of concrete methods of tangible representation. Such investigation must also remain alert to possibilities for *manipulation*, since the possibilities that are identified for tangible display and tangible manipulation will together enable as well as delimit the designs than can ultimately realize improved instrumental control.

### 1.3 Approach: force output to the fingerpad

When implementing tangible display to dynamically represent separate grains generated by algorithmic processes, this will first require choices in anatomical location and means of delivery. The hands can be considered as the most versatile parts of the human body for sensing and manipulating the immediate tangible surroundings. For fine sensing and manipulation, the fingertips especially are used as the areas of contact, having the highest spatial resolution in the cutaneous (skin-based) sense of touch across the hand [5]. Such fingertip contact will often involve the fingerpad skin areas, which have been used for the instrumental control of musical sound over tens of millennia, e.g. to close the holes of flutes [1], pluck sounding strings, press keyboard keys, etc.

Here, we will consider flat, stiff surfaces, put in contact with the fingerpads to apply forces, controlled over time, to the fingers. In general, this can result not only in cutaneous but also in kinesthetic sensations of touch involving finger movement. We will use two general interfaces for touch in instrumental control of musical sound, which have been described elsewhere: the cyclotactor ("CT") [2] and the kinetic surface friction renderer ("KSFR") [3]. In the CT, the flat surface is attached to the fingerpad using a strap. Voluntary fingerpad movements are intended to happen only perpendicularly to the fingerpad's surface. In the KSFR, the flat surface is pressed down upon. Voluntary fingerpad movements are then limited to happen in parallel to the fingerpad's surface. This is shown in Figure 1, where the different types of intentional movement and applied force in the two interfaces are described and illustrated in more detail.

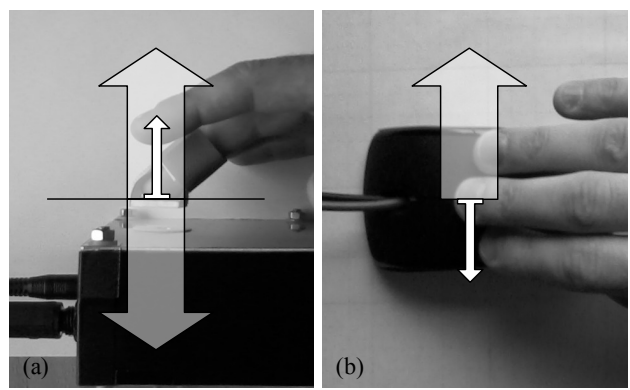

**Figure 1. Fingerpad movements and applied forces in the two interfaces used.** The large transparent arrows indicate directions of intentional movement; the small opaque arrows indicate the direction of the controlled force components that are applied.

(a) The CT setup: intentional movements are performed perpendicularly to the fingerpad surface. Regardless of whether the fingerpad is intentionally moved up, down or held still, the direction of applied forces will (here) be perpendicularly against the fingerpad.

(b) The KSFR setup: intentional movements are performed parallel to the fingerpad surface. Forces are applied during movement, and are opposed to the direction of movement (only one case is shown).

## 2. PILOT EXPERIMENT
### 2.1 Overview

To investigate possibilities for tangible display of granular synthesis using the two general methods of force delivery to the fingerpad, an informal pilot experiment was conducted. In it, both force magnitude and headphone sound output were controlled over time. Both were determined by the same variable, on/off master block impulse signal. This master block impulse controlled sound output by modulating a sine wave signal of a relatively high frequency, allowing the signal to retain pitch more easily for shorter impulse durations. The master block impulse controlled force output in a similar way, by modulating either a sine wave signal or a level maximum amplitude signal. Here, the sine wave used had a frequency of 250 Hz, placing it within the frequency region where the vibrational sensitivity of mechanoreceptors is highest [6]. To help create the impression of a single "grain event" occurring on both channels, millisecond latencies were adjusted so that the patterns in sound and force output would temporally coincide as much as possible. As cannot be seen in Figure 1, a single finger was used to contact the stiff surface of the KSFR during intentional movement. Tables 1 and 2 describe the experimental parameter values that were kept constant and those that were varied, respectively.

**Table 1. Experimental parameters kept constant.**

| interval between successive grain event onsets: | 1.00 s |
|---|---|
| sound block impulse maximum amplitude: | constant |
| sound carrier signal sine frequency: | 4000 Hz |
| baseline force level: | 0.14 N |

**Table 2. Experimental parameters that were varied.**

| interface: | CT / KSFR |
|---|---|
| master block impulse duration: | 100 / 50 / 10 / 1 ms |
| force block impulse max. amplitude: | 1.00 / 0.72 / 0.43 N |
| modulated force signal: | constant / 250 Hz sine |

## 2.2 Results

In the CT interface, both with and without headphone output, the non-vibratory force impulses generated seemed clearly perceivable for all of the impulse durations tested. The differences in these impulse durations, as well as the differences in amplitude at given impulse durations also seemed clearly perceivable. Of the vibratory force impulses, only the durations above 1 ms were considered, since only these would fit at least one vibration wave cycle (of duration 4 ms). For these durations, both the differences in duration and in amplitude seemed clearly perceivable. The type of sensation seemed to change with duration: at 100 and 50 ms, impulses seemed to give an impression of vibration, while at 10 ms, this changed to a pulsed sensation that seemed less distinct when compared to a non-vibratory impulse of the same duration.

In the KSFR interface, force impulses of duration 1 ms could not be considered due to a technical issue: at this duration, a mechanical effect in the housing of the device resulted in the perception of forces in the fingerpad also when it was being held still. This made it ambiguous whether apparently weak forces applied in parallel to the fingerpad surface during movement were being felt separately of this, or not. At the remaining durations of 10 ms and higher, however, these forces seemed well distinguishable both for the non-vibratory and vibratory force impulses. Also, both the differences in duration and in amplitude at each duration seemed perceivable. Here too, the type of sensation seemed to change with duration for vibratory force impulses: at 50 and 100 ms these gave an impression of vibration, while at 10 ms this again changed to a pulsed sensation, not unlike that produced by a non-vibratory impulse of the same duration.

In both interfaces, force impulses with larger amplitudes and durations were clearly able to influence position and speed input. In the CT interface, this resulted in vertical displacements of the fingerpad; in the KSFR interface, it resulted in the slowing down of intentional movements. Figure 2 below shows an example recording of output by the CT interface, at the 'microtouch' end of the temporal scale.

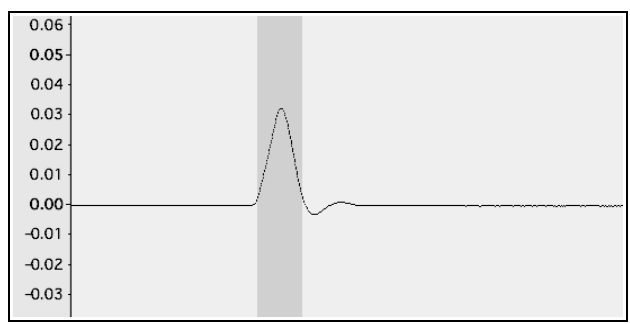

**Figure 2. Microtouch output**. Linear magnetic field strength recording, made during the application of a perceived force impulse by the CT interface. The grey vertical bar represents a duration of 1.0 ms.

## 3. DISCUSSION AND FUTURE WORK

The results of the informal pilot experiment indicate that the perpendicular and parallel methods of force delivery to the fingerpad as implemented in the CT and KSFR interfaces can be used for the tangible display of separate grain events, meaningfully operating in the same time scale that underlies the granular synthesis of musical sound. Specifically, it seems that variations in the amplitude and duration of applied force impulses could be used to dynamically mirror or display aspects of grain generation, starting from the level of separate grains. For longer impulse durations, it seems vibratory force could add an additional dimension to such display.

Of the two interfaces tested, it seems the CT is currently somewhat better positioned to display fine detail in applied forces developing over time.

It may seem self-evident that for successfully improved forms of instrumental control of granular synthesis to be realized, the musical sound output and tangible display of systems should be developed in close tandem. This seems the more so since signals to one sense can influence the perception of other signals to other senses in many ways: for example, vibrotactile stimulation influences the sensation of hearing a tone [11].

*Can grains become manipulanda?* One way towards this suggested by the results from the pilot experiment seems to be using the changes in displacement and velocity input caused by force impulse output – as net displacements will be the result of forces applied by both the interface and the user.

For this reason and the reasons stated at the beginning of this section, based on the fundamental motivating factors discussed in the introduction, it seems that the methods of dynamically applying force to the fingerpad presented here should be further investigated for their potential to enable new and appropriate forms of tangible display and manipulation for the instrumental control of granular musical sound.